\documentclass[%
 reprint,
superscriptaddress,
 amsmath,amssymb,
 aps,
]{revtex4-1}

\usepackage{graphicx}
\usepackage{subfigure}
\usepackage{dcolumn}
\usepackage{bm}
\usepackage{xcolor}
\usepackage{lineno}

\def\nin{\noindent} 
\def\beq{\begin{equation}}
\def\eeq{\end{equation}}

\begin{document}

\title{Self-organisation - the underlying principle and a general formalism}

\author{Raphael Blumenfeld}%
 \email{rbb11@cam.ac.uk}
\homepage{https://rafi.blumenfeld.co.uk}
\affiliation{
Gonville \& Caius College, University of Cambridge, Trinity St., Cambridge CB2 1TA, UK}

\date{January 25, 2026}
 
\begin{abstract}
It is proposed that self-organisation (SO) in non-equilibrium systems is governed by a general principle: it emerges when a minute subset of system configurations are exceptionally stable and long-lived to survive the noise generated by the driving and environmental constraints. 
Guided by this principle, a statistical mechanics-like model is formulated for general SO and its application is illustrated for two example systems: self-organised steady states of quasi-statically driven granular systems in two dimensions and crowd laning, for which illustrative explicit results are derived.  In this formalism, maximising a survivability function of the exceptionally few stable configurations is the equivalent of minimising the free energy in traditional statistical mechanics. 
Parallels with equilibrium statistical mechanics and differences from it are discussed, which provides useful insight to assist in modelling SO in general out-of-equilibrium systems. Similarities and differences between SO in passive and biological systems are also pointed out, suggesting potential extension of this approach in this direction, albeit to very simple systems.

\end{abstract}


\maketitle


\nin I. \underline{{\bf Introduction}} \\
The term self-organisation (SO) refers usually to spontaneous emergence of large-scale ordered patterns or symmetries in driven disordered out-of-equilibrium systems. 
This phenomenon is ubiquitous, from formation of structural patterns in inanimate matter, such as snowflakes, dune ripples, and some cloud formations, to temporal manifestations and evolution of living organisms. 
SO, often associated with complexity and pattern formation, emerges as a result of local interactions between the individual system constituents and is not externally engineered. While this description also applies to traditional phase transitions in thermal systems, in which the competition is between energy and entropy, it is much more general, manifesting in physical, chemical, biological, active, and social systems~\cite{SOR1,SOR2,SOR3,SOR4,SOR5}. \\
Most studies of SO have been usually focused on specific contexts and in many case are descriptive. 
Attempts at finding a general principle underpinning SO are few and they are mainly based on energy considerations. For example, it has been proposed that SO emerges in steady states by selection of microstates that minimise energy-dissipation in driven systems~\cite{Cyetal21,Keetal23}.  These models propose that noise 'rattles' the system and, as it is bumped from microstate to microstate, it gets trapped in special microstates where the noise is very low. Yet, the physical mechanism for the selective low rattling in those microstates is unclear. A physical mechanism is suggested in~\cite{Keetal23}, based on inability to escape energy potential wells. However, it is unclear if this explains other energy-dissipating systems and how to extend it to SO in athermal systems, where thermal noise is negligible compared with noise from other sources, especially when energy-based considerations plays little to no role in determining the self-organised state.
Observations of SO in the latter systems suggest that a more general principle is in play, which applies to all systems, however driven. 
Motivated by this idea, the main aim here is to state the fundamental principle that governs SO phenomena in general and, based on it, propose a formalism to determine their characteristics.
The approach is analogous to statistical mechanics, a method that has been applied in recent decades well beyond thermal equilibrium, e.g.~\cite{EdOa89a,EdOa89b,Ga99,Vietal99,Ya09,Taetal24}.
Like traditional statistical mechanics, the formalism can predict the characteristics of the long-term self-organised steady states in general - it does not model the emergence SO from specific driving dynamics.
A significant advantage of this approach is that the parallels between the proposed formalism and traditional statistical mechanics offer insight into better modelling of SO in non-equilibrium systems.

For didactic reasons, the principle is first illustrated in the context of a specific example - driven granular systems, and its generalisation is discussed. 
For the same reason, the approach is also formulated initially in the context of this system. While the formalism is general, to demonstrate its its use in specific systems, some assumptions must be made and these are stated, under which expectation values and an equivalent of an equation of state are derived explicitly.  
The method is then formulated in general and discussed. 
This is followed by an illustration of the application to another phenomenon - a simple model of crowd laning SO. 
The examples are intended as demonstration of the use of the formalism for SO that is not determined by energy considerations, rather than as better modelling of those specific systems.  
A discussion follows of potential applications of the formalism to simple biological systems. 
The paper concludes with a summary, a discussion of similarities and differences between this method and equilibrium thermal systems, and possible future work.  \\

\nin II. \underline{{\bf SO in 2D granular systems and the principle}} \\
Recent observations of SO in experiments and simulations of 2D granular systems provide the key to identify the principle governing this phenomenon~\cite{MaBl14,Waetal20,Suetal25,MaBl17,Suetal20,MaBl21,Jietal25}. Driven quasi-statically out of equilibrium by various methods and for a wide range of parameters, it has been found that: many measurable quantities appeared to collapse to master curves~\cite{MaBl14}, structural transitions converge to detailed-balance-like steady states~\cite{Waetal20,Suetal25}, and local stress and structural configurations self-organise cooperatively~\cite{MaBl17,Suetal20,MaBl21,Jietal25}. All these signatures involved the cells, which are the smallest possible voids surrounded by particles in contact. The latter is very significant; when cells were approximated as ellipses, their orientations were strongly correlated with the local principal stress axes. As cell orientations were isotropic and independent of the boundary conditions, this was strong evidence of a cooperative stress-structure SO. The physics behind this correlation is not difficult to understand. The driving generates noise, causing the internal stress field to fluctuate in space and time. A cell that cannot support the local stress disintegrates, i.e., at least one of the contacts of the particles surrounding it breaks. This causes the stress around it to reconfigure and stabilise when a new cell is formed. The stablest configuration corresponds to the cell's ellipse aligning with the largest principal local stress (think of resistance to pressing on an egg along its long or short axes). 
Being stable, these cell shape-stress configurations are sufficiently long-lived to be observed consistently, i.e., the system self-organises into them. This sheds light on a general principle that governs SO -  a minute subset of configurations dominates observations because of their higher stability and longer survivability. Extending this interpretation to a general principle is discussed below.\\

\nin III. \underline{{\bf SO statistical mechanics in granular systems}} \\
As occurrence probabilities of long-term configurations are at the core of SO, a natural modelling approach is statistical mechanics - a method that derives system-wide properties on the basis of such probabilities. 
For the granular systems, cells are chosen as quasi-particles because all the aforementioned signatures involved these structural descriptors. Being defined by the contacts of particles around them, cells have the advantage of sensitivity to the local forces, which are transmitted via the contacts. It is this sensitivity that gives rise to the strong correlations between cell shapes and local stresses~\cite{Jietal25}. Descriptions based on Voronoi tessellations, for example, smear this correlation because they are based on particle proximity rather than contacts. The actual grains are also less useful as quasi-particles because their shapes are fixed and their characteristics are not as sensitive to local stress fluctuations.
A cell surrounded by $k$ particles in contact is a cell of order $k$, or $k$-cell. 
 The cell order distribution across the system is discrete, with $Q_k$ ($k=3, 4, ..., K$) the fractions of $k$-cells out of all $N_c$ cells. 
Cell stability depends on two 'attributes': its shape, approximated as an ellipse~\cite{Jietal25}, and its stress. A cell ellipse has three degrees of freedom (DOFs): the major and minor axes lengths, $a_c$ and $b_c$, respectively, and the orientation of the major axis relative to a fixed frame of reference, $\theta_c$. 
Cell stresses are defined, using an appropriately weighted sum over the stresses of the particles surrounding it~\cite{Bletal25}, as detailed in the supplemental material~\cite{SM}. In 2D, the cell stress also has three DOFs: the large and small principal stresses, $\sigma_{c1}$ and $\sigma_{c2}$, respectively, and the orientation of the large one relative to the same fixed frame of reference, $\phi_c$. 

At the heart of the statistical mechanics formalism is a partition function, which accounts for the occurrence probability of each system-wide configuration. In the proposed approach, the more stable the cell's shape and stress the higher is its occurrence probability and the longer it survives. The smaller the difference between the its ellipse axes lengths and its principal stresses the stabler it is. 
A standard stress stability measure is what is known as the stress ratio~\cite{VMi13}
\beq
h_c\equiv \frac{\sigma_{1,c} - \sigma_{2,c}}{\sigma_{1,c} + \sigma_{2,c}} \ .
\label{StabilityMeasures}
\eeq
Any model for the shape stability can be used and, for illustration purposes, assume a similar form,
\beq
f_c\equiv \frac{a_c - b_c}{a_c + b_c}  \ .
\label{StabilityMeasures}
\eeq
The rationale behind this definition is that long narrow cells, for which $f_c\to1$, are very unstable and vice versa.
Next, a survivability function is required, whose value determines the configurations occurrence probabilities. It should depend on the stability measures, on the DOFs, and on constraints on the system.  Its main aim is to quantify the stability, and thence survivability, of system-wide configurations. 
The survivability function should take into considerations the observed function collapses and the shape-stress correlations as coupling between $\theta_c$ and $\phi_c$ at the cell level. 
In principle, it could also include correlations between the DOFs of nearest-neighbour cells, although searching for such correlations experimentally revealed none~\cite{Waetal20,Suetal25}. Making the survivability function decrease with stability would make it analogous to the free energy in thermal systems. 
It is natural to use a Boltzmann factor-like exponential to weigh the configurations occurrence probabilities. This is because the stability is an extensive quantity - the overall stability is the sum over the stability measures of the individual cells. The exponential is the only function that accommodates extensivity in that the occurrence probability of a system of independent subsystems is the product of their individual probabilities and its extensive properties are the sum of the individual ones. 
Finally, the partition function should also be grand-canonical because the number of cells, $N_c$, fluctuates.  
 All these specifications are satisfied by 
 \begin{align}
Z=\prod\limits_{k=1}^K \left[\left(\sum_{c\in k}^{Q_kN_c} e^{-F_c} \right) e^{-\mu Q_kN_c}\right] \ ,
\label{Z0}
\end{align} 
with a cell survivability function
\begin{align}
F_c = &\ J_1 f_c(k) + J_2 h_c(k)   \notag \\
&+ J_3 \left[\theta_c(k) - \phi_c(k)\right]^2  + J_{c'c} \psi_{c'c}^2 \ .
\label{Fc}
\end{align}
In this expression, $\mu$, $J_1$, $J_2$, and $J_3$ are Lagrange multipliers that reflect the effect of the dynamics-generated noise on, respectively, the number of cells, the cell shape stability, the cell stress stability, and the cell-shape orientational correlations. The choice of a square form for the third term is for simplicity, because it enables analytical calculations, and in the absence of any other data on it. In general, any functional form may be used, as long as it increases with the difference between the angles such that the survivability function decreases monotonically with it.
The last term represents the cell-cell interaction, with $\psi_{c'c}=\left(\theta_c-\theta_{c'}\right)^2$ the difference between the orientations of neighboring cells $c$ and $c'$. More interaction terms between other DOFs of neighbouring cells can be added to $F_c$, if necessary.

As the DOFs are continuous, it makes sense to convert the summation in (\ref{Z0}) into integrals of `densities of states' over the phase space of DOFs. 
Denoting the cell DOFs by $\left\{X\right\}=\left\{a, b, \theta, \sigma_{1}, \sigma_{2}, \phi  \right\}$, 
the densities of states provide the occurrence probabilities of the DOFs and/or the cells stability measures. 
For the 2D granular systems, some of these distributions are known from modelling and observations. 
Specifically, $\theta_c$ is isotropic on the cell level and $\phi_c$ is strongly correlated with it~\cite{Jietal25}. Thus, these two variable can be replaced by one variable, $\Delta\equiv\theta - \phi$. This variable is normally distributed almost identically around zero for all $k$, $P(\Delta)\sim e^{-\Delta^2/\left(2\delta^2\right)}$. 
In principle, the distributions of the stability measures $h$ and $f$ can be obtained from the distributions of the DOFs on which they depend. However, it has been shown that, in these systems, the conditional distributions of $h_c$ given $k$, $P_h(h|k)$, when scaled by the mean for each cell order, $\bar{h}(k)$, collapse onto a unique Weibull form, $W\left[\hat{h}=h/\bar{h}(k)\right]$~\cite{We51}, that is independent of $k$~\cite{Jietal25}. 
A model for this collapse is detailed in the supplemental material~\cite{SM}, predicting almost exactly the experimental observations,
\beq
W(\hat{h}) =  g_0\hat{h} e^{-\frac{g_0}{2}\hat{h}^2} \ ,
\label{DifEq}
\eeq
with $g_0$ a parameter determining transition probabilities between states. This PDF is used in the following calculations.
Following the observations in~\cite{Waetal20,Suetal25} and for simplicity, cell-cell correlations are neglected next, $J_{c'c}=0$, and the formalism reduces to an `independent cells model'. 
For the illustration here, the conditional density of states of $f(k)$ are assumed to collapse onto a master form of the same form as of $\hat{h}$. An explicit calculation of the partition function, detailed in the supplemental material~\cite{SM}, yields 
\begin{align}
\ln{Z}=\Bigg\{
&\sum\limits_{k=1}^K Q_k \ln\left[\langle e^{-J_1\bar{f}(k)}\rangle \langle e^{-J_2\bar{h}(k)}\rangle\right] \notag \\
&- \mu - \frac{1}{2}\ln{\left(1+2\delta^2 J_3\right)}
\Bigg\}N_c \ ,
\label{Z1}
\end{align}
with the angular brackets denoting averages over all the DOFs. This expression can be used to obtain expectation values of system-wide measurable quantities as functions of the noise parameters and coupling constants in the survivability function,
\begin{align}
\langle G\rangle = \frac{1}{Z}\Bigg\{&\prod\limits_{k=1}^K e^{-\mu} \int G\left(\{ X\}\right) e^{-J_1f-J_2h-J_3(\theta-\phi)^2} \notag \\ 
&P\left(\{ X\}\right)d^6\left\{X\right\} 
\Bigg\}^{Q_k N_c} \ .
\label{Z1}
\end{align}
For example, the expectation values $\langle \hat{f}\rangle$ and $\langle \hat{h}\rangle$ are
\begin{align}
\langle \hat{f}\rangle &= \sum\limits_{k=3}^K Q_k\bar{f}(k) = -\frac{\partial\ln{Z}}{\partial J_1}  \\ 
\langle \hat{h}\rangle &= \sum\limits_{k=3}^K Q_k\bar{h}(k) = -\frac{\partial\ln{Z}}{\partial J_2}  \ .
\end{align}
Explicit forms of these, based on the above assumptions, are derived in detail in the supplemental material~\cite{SM} and are too cumbersome to include here. 
Also derived in the supplemental material is the (cumbersome) relation
\begin{align}
\langle \hat{f}\rangle = \prod\limits_{k=1}^K 
\left\{\left[\frac{u' - v'\chi'}{u - v\chi}\frac{w -2u\chi + v\chi^2}{w' -2u'\chi' + v'\chi'^2} \right]\right\}^{Q_k} \langle \hat{h}\rangle \ ,
\label{EoS}
\end{align}
with all the quantities in this expression defined there. As $\langle \hat{f}\rangle$ and $\langle \hat{h}\rangle$ are system-wide characteristics, (\ref{EoS}) is the equivalent of an equation of state. 
Significantly, this is a relation between properties of the stress and the structure.  \\

The extension of this formalism to three-dimensional (3D) granular SO is conceptually straightforward. As in 2D, cells are the smallest polyhedral voids surrounded by particles in contact. They can be approximated as ellipsoids and their stresses defined as weighted sum of the stresses of their surrounding particles. This would give 10 DOFs per cell (three axes and two angles for each of the shape and stress ellipsoids). Similar ellipsoid and stress stability measures can be defined and $F_c$ can be a straightforward 3D version of (\ref{Fc}). 
This, in principle, provides the partition function. However, its evaluation may be more difficult in 3D because of the more complicated cells classification into families.  
While the 2D cell polygons can be classified only by their order, classification of a 3D polyhedral cell requires more parameters: the number of its faces and the order of each polygonal face. Specifically, a family of cell polyhedra is classified by its $n_p$ vertices (contacts), $k_3$ triangular faces, $k_4$ rectangular faces, and so on up to the highest-order faces, $k_t$. Members of the same family have an identical series of such numbers.  
Nevertheless, this complication is a quantitative, rather than qualitative. It just means that the product over $k$ in (\ref{Z0}) needs to be replaced by products over all the cell parameters, but the integrand is the same. 
This extension is under investigation currently, but it is downstream from the aim here. \\

\nin IV. \underline{{\bf The principle and its generalisation}} \\
The emergence of SO in granular systems suggests a general principle. SO corresponds to minimising entropy in processes where a minute subset of configurations have a much higher occurrence probabilities than the overwhelmingly larger number of configurations. This interpretation is supported by further numerical and experimental observations in disordered disc assemblies. It was found that the mean area of $k$-cells, when normalised by the area of the regular $k$-polygon of the same edge length, first decreases with $k$ and then increases again~\cite{MaBl17,Suetal20}. The initial decrease is the result of an increasing number of elongated cell configurations, which are stable at low cell orders but become increasingly unstable as $k$ increases. It follows that the effect of stress on the structure and cell stability limits configurational entropy. 
 
This principle holds beyond granular systems. Driving a disordered system generates noise even if thermal fluctuations are negligible. Such systems are also constrained by the environment through boundary conditions and interactions with other dynamic systems, as well as internally by the nature of interactions between their individual components.
SO emerges when only a minute subset of system configurations survives the noise and constraints sufficiently long to dominate experimental observations. As the bulk of configurations disintegrate much faster, they are not typically observed. 
Different driving conditions lead to different noise characteristics and to different SO characteristics, as it is the noise that determines which configurations survive the longest. \\

\nin V. \underline{{\bf A general modelling framework}} \\
Extending the modelling formalism to more general systems is as follows.  
(1) define quasi-particles (e.g., the cells); 
(2) identify the quasi-particles' DOFs (the cell ellipse axes lengths and orientations, principal stresses and orientation); 
(3) identify quasi-particle stability measures that depend on the DOFs and whose values decrease with increasing quasi-particle stability (e.g., the cell stresses and shapes); 
(4) construct a survivability function that decreases monotonically as quasi-particles' stability increases (e.g., $F_c$ in eq. (\ref{Fc})).  The value of the survivability function is determined by each system-wide configuration's stability and it determines the configuration's occurrence probability and lifetime. \\
Consider then a system of $N$ elements, structural or otherwise, which show signatures of SO, and define these as quasi-particles, $q=1,2,...,N$. Associated with each, are $M$ properties, indexed $m=1, 2, ..., M$. These properties self-organise locally for every quasi-particle across the system. Property $m$ of quasi-particle $q$ depends on $A_m$ of its DOFs, indexed $a_{q1}^{m}, a_{q2}^{m},...,a_{qA_m}^m$. 
The modelling of this system requires identification of $f^m_{q}$ stability measure of each property $m$ of quasi-particle $q$. 
The quasi-particle survivability function, $F_q$, depends on these measure, as well as on other DOFs. It can also involve interactions between DOFs of the same quasi-particles, as well as different ones. For example, a generalisation of (\ref{Fc}) would be
\begin{align}
F_q = &\sum\limits_{m=1}^M\sum\limits_{j=1}^{A_m}J^{m}_{j}\left(a^m_{qj}-a^m_{qj,0}\right)^2  +  \notag \\
&\sum\limits_{q'}\sum\limits_{m,n=1}^M\sum\limits_{j=1}^{A_m}\sum\limits_{k=1}^{A_n}\lambda^{m,n}_{qj,q'k}\left(a^m_{qj}-a^n_{q'k}\right)^2  \ ,
\end{align}
where: $a^m_{qj,0}$ is the mean value of $a^m_{qj}$ at perfect SO; $J^m_j$ a Lagrange multiplier-like parameter that weights the significance of such a fluctuation in terms of occurrence probability; $q'$ are all the quasi-particles that interact with $q$, e.g., its nearest neighbours; and $\lambda^{m,n}_{qj,q'k}$ the coupling between DOF $a^m_{qj}$ and $a^n_{q'k}$. 
The phase space is spanned by all the DOFs, whose number is 
$${\cal{N}}_{DoF}=\sum\limits_{q=1}^N\left\{\sum_{m=1}^M\left(\sum_{j=1}^{A_m}a_{qj}^m\right)\right\} \ .$$
As the DOFs are usually continuous, the general canonical partition function is \\
\begin{align}
Z = \prod\limits_{q=1}^N \left\{\int e^{-F_q-\mu} \left[\prod\limits_{m=1}^M\left(\prod\limits_{j=1}^{A_m} d a^m_j\right)\right]\right\}  \ .
\label{ZGen}
\end{align}

\nin VI. \underline{{\bf Example application - Laning SO}} \\
To illustrate the application of the formalism, consider the phenomenon of self-organised laning~\cite{He71,HeMo95,Chetal00}.  
Laning is a spontaneous formation of lanes as agents (pedestrians, cars, etc.) move in crowded spaces. It is the result of agents aiming to move in a particular direction or reach a particular destination without bumping too often and too violently into others. While collisions, and their avoidance, are local, ordered system-wide laning often emerges. This SO makes the flow of agents more efficient, e.g., by minimising the time spent in the crowd before reaching the destination. 
The global lanes pattern consists a minute fraction of all the possible ways that the agents can move.
As the aim is to illustrate the formalism, rather than improve existing models of this phenomenon, simplifications are made for calculating the partition function proposed below.

\begin{figure}[!h]
 \includegraphics[width=1.0\linewidth]{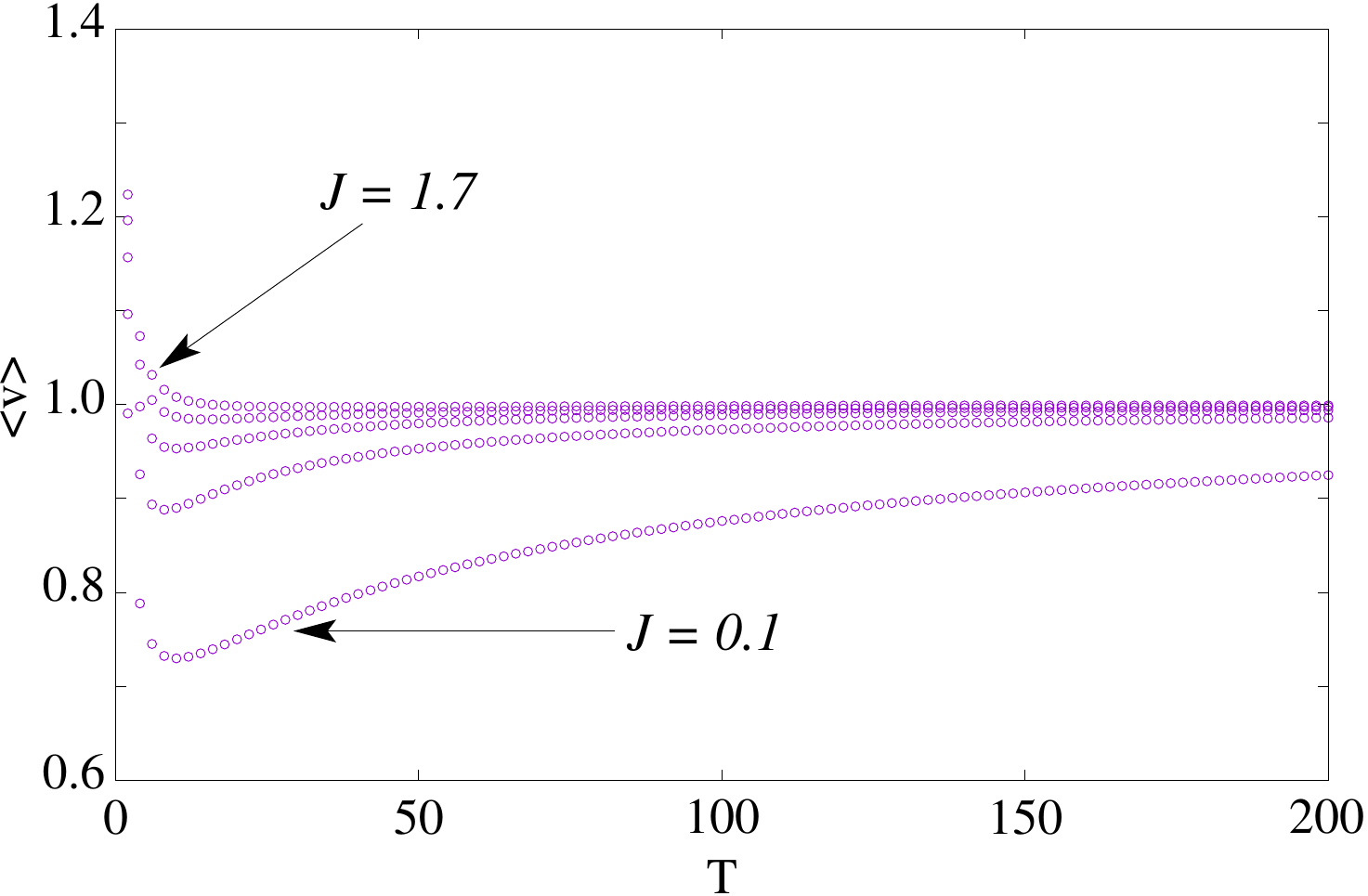}
 \caption{The mean speed of the two lanes, derived in the mean field approximation, as a function of the noise parameter, $T=1/\lambda$, for several values of the bias parameter $J$. A transition from a monotonic decay to a minimum at some $T_c$ emerges as $J$ decreases below $J_c\approx1.33$.}
 \label{MeanSpeed2}
\end{figure}

In a simple version of this problem, $N$ agents move in one direction, say $y$, along a long strip of width that allows only few agents to move side by side. They can move at one of two speeds $v_1$ and $v_2>v_1$ and at any angle relative to the forward direction, $-\pi/2<\theta<\pi/2$. Moving at $\theta=0$ is fastest, but can be hampered by collisions with more slowly-moving agents. Attempting to overtake those by moving sideways may incur collisions with other agents. 
Here, the agents are the quasi-particles and $q=1, 2, ..., N$ and each agent has two DOFs - the speed and direction of motion.  
An agent's most sustainable mode of progress is by avoiding collisions, achievable with the stability measure,
\beq
f_q = \lambda\sum\limits_{q' \Rightarrow q}\left(\vec{v}_q-\vec{v}_{q'}\right)^2  \ .
\eeq
where $q' \Rightarrow q$ denotes the indices of the agents nearest to $q$ (identifiable, e.g., using Voronoi tessellation at each configuration). 
The objective of reaching the destination efficiently is a constraint that can be quantified in the survivability function by encouraging motion in the forward direction. A natural survivability function is
\beq
F_q = J \tan^2{\theta_q} + \lambda\sum\limits_{q' \Rightarrow q}\left[\left(\vec{v}_q-\vec{v}_{q'}\right)^2 + \left(v_{qy}-u_{max}\right)^2\right] \ .
\eeq 
Here, $J$ biases motion forward and the last term encourages motion forward as quickly as possible.  Further constraints and specifications can be introduced by adding appropriate terms in $F_q$. 
The partition function is  
\begin{align}
Z = \int  e^{-\sum\limits_{q=1}^N F_q} \notag \times \prod\limits_{q=1}^N \Big\{P\left(v_q,\theta_q\right)d v_q d \theta_q\Big\} \ ,
\end{align}
with, say, $P\left(v_q,\theta_q\right)=\left[p\delta(v-v_1)+(1-p)\delta(v-v_2)\right]/\pi$ the density of states. 
It is clear that the interplay between $J$ and $\lambda$ determines that the stable system-wide configurations are the small subset of those that approximate two lanes, one moving at $v_1$ and the other at $v_2$. 
The partition function allows us to predict the characteristics of these configurations and it has been evaluated explicitly in the supplemental material~\cite{SM} in an effective medium approximation for a somewhat more general case (eq. (27) there).  Using it, the mean lane speed has been calculated
\begin{align}
\langle v\rangle = \frac{\sum\limits_{m=1}^2 \frac{S_m}{\sqrt{\xi_m}} v_m e^{- 2\lambda K v_m\left(v_m-u_y-u_{max}\right) + \xi_m C_m^2}}
{\sum\limits_{m=1}^2 \frac{S_m}{\sqrt{\xi_m}}  e^{- 2\lambda K v_m\left(v_m-u_y-u_{max}\right) + \xi_m C_m^2}} \ ,
\end{align}
with the variables defined in the supplemental material. A plot of this expectation value is shown in Fig.~\ref{MeanSpeed2}, in which $T=1/\lambda$ is used as a scaled measure of the noise, in analogy to temperature in thermal systems. In very noisy systems, i.e., with many collisions between agents, the mean progress speed converges to the average of the lanes' speeds, $\langle v\rangle\to\left(v_1+v_2\right)/2$. At very low noise, $T\to0$, the speed is dominated by the fast movers, $\langle v\rangle\to v_2>v_1$. Since collisions rate (=noise) increases with agents density, this behaviour is consistent with the results in~\cite{Heetal06,Moetal12}. This model also reveals an intriguing phase-like transition: while at high values of the bias parameter, $J$, $\langle v\rangle$ decreases monotonically with $T$, as $J$ decreases, a minimum merges at some $T_c$. This demonstrates that the approach presented here can yield rich behaviour even for simple models. Analysing this transition in more detail is downstream from the main thrust of this paper. \\

\nin VII. \underline{{\bf Potential relevance to biological SO}} \\
Both the general principle and statistical mechanics method, proposed here, can apply to biological systems: proteins, cells, organisms, and entire species. 
In these contexts, the principle translates straightforwardly to ‘the fittest survive’. The formalism can also be extended by inspecting the equivalence between inanimate and biological systems. An organism's survival is based on functionalities and skills that enhance its 'longevity' and, consequently, that of the species. The equivalent of noise and fluctuations are factors that could shorten longevity and survivability, e.g., genetic variability, intra-species competition for reproduction, availability of resources, existence of predators, environmental cataclysms, etc.. An individual biological system is the equivalent of a quasi-particle, whose configuration is equivalent to the set of biologically inherited and adapted functionalities and skills. Better such skills improve 'fitness' and survivability equivalently to better stability measures that increase the lifetime of a configuration to withstand the disruptive noise. \\
The main difference is that biological systems are active and can adapt, evolve, and develop feedback mechanisms to modify specific functionalities. This makes it possible for these systems to respond to noise in ways that passive systems cannot. In other words, they modify stability measures, e.g., by honing existing skills or developing feedback mechanisms, to improve configuration survivability against fluctuations. Active systems can also improve the survivability function, by adding skills, which is equivalent to adding terms in the survivability function, or by changing the environment, which is equivalent to changing the noise parameters. Biology is much richer because the noise , e.g., predators and resources, is also adaptive and changing in response. Biological reproduction also has an equivalent in passive systems - a line of descendents can be regarded as a sequence of the stable configurations in self-organised steady states. 
In short, as the objective of adaptation, evolution, reproduction, and feedback mechanisms is always to increase survivability of individuals and species against fluctuations and constraints, they can be described, in principle, by the proposed survivability statistical mechanics.  \\

Nevertheless, modelling complex organisms with the proposed formalism requires taking including in the survivability function all the noise sources, the functionalities, the relevant skills, and so on. Therefore, the proposed formalism is likely to be applicable to simple biological systems. It should be emphasised, however, that this is a quantitative difficulty rather than a conceptual.  \\

\nin VIII. \underline{{\bf Concluding discussion}} \\
A general principle, which governs SO in out-of-equilibrium driven disordered systems, has been formulated. It states that SO emerges when a minute subset of system configurations are exceptionally stable and survive the noise, generated by the driving and other constraints, sufficiently long to dominate  observations. The apparent reduction in entropy is because the overwhelming majority of theoretically possible configurations are dismantled by the noise and are too short-lived on observation timescales. 
This principle extends an earlier principle, based on energy considerations~\cite{Cyetal21,Keetal23}, to SO phenomena determined by general mechanisms~\cite{Jietal25,HeMo95,Chetal00}. \\
 
Based on  this principle, a statistical mechanical-like formalism has been proposed to model the characteristics of the long-term self-organised state. It  involves identification of quasi-particles, their DOFs, stability measures, and a survivability function, whose value decreases with the stability measures. 
A partition function has been constructed, in which the occurrence probabilities of system configurations are determined by Boltzmann-like exponentials to accommodate extensivity. The SO is determined by the high occurrence probability of configurations that minimise locally the survivability function.  \\

The formalism makes possible derivation of relevant expectation values of the self-organised state, equivalents of equations of state, and phase-like transitions. This predictive power has been demonstrated for two examples: a cooperative SO in 2D granular experiments and the laning phenomenon. In the former, expectation values of local structural and stress stability measures have been derived, as well as the equivalent of an equation of state. The extension to three-dimensional granular systems has also been outlined and discussed in detail. 
In the second example, a simple model has been constructed for the laning phenomenon, which enabled a derivation of the mean lane speed. Surprisingly, the method provides a prediction of a non-trivial phase-like transition as a function of noise and bias parameters. In both these examples, the SO is not determined by energy considerations. While hardly any literature exists about relations between stability measures and long-term SO characteristics, as defined here, these are straightforward to obtain in simulations and experiments, which would help test some of the explicit results derived here. 
Nevertheless, it should be emphasised that the examples presented here come mainly to illustrate the use of the formalism and the way that it used to derive steady state characteristics. Whether these characteristics describe correctly the modelled phenomena depends crucially on the models used for the survivability function. This is no different from statistical mechanics in thermal systems, in which the choice of the relevant Hamiltonian is essential for describing a system correctly.  \\

There are parallels between this formulation and its analogue in thermal systems, in which the microstates are the system-wide configurations, particles' energies and potentials are used as stability measures, and the thermal survivability function is the free energy. Like the free energy, the survivability function depends on the quasi-particles' DOFs and its lowering increases a microstate's occurrence probability and hence its survivability. Following this rationale, some equilibrium macrostates can also be viewed as self-organised. An example is solidification under temperature lowering, where the overwhelming number of microstates with very energetic particles are strongly unstable and only the minute ordered microstates survive to be observed. The lowering of temperature is analogous to increasing $J_i$ in (\ref{Fc}). \\
There are, however, differences. The energy in thermal systems is a constant of the motion, from which dynamic equations of motion of the DOFs can be derived. The survivability function only describes stability in long-term out-of-equilibrium self-organised steady states and, as far as this author knows, cannot play such a role. 
Additionally, thermal systems incur indigenous thermal fluctuations while in athermal systems it is the driving that gives rise to noise.  
Nevertheless, the similarities to traditional statistical mechanics are useful as guidelines for constructing general models of SO. \\

The principle and the modelling approach are applicable, in principle, to SO of biological systems. In this context, the principle translates to ‘the fittest survive’, i.e., the more stable the biological system the more fit it is to survive its noisy environment. The equivalence of stability measures and survivability functions between SO in such systems and in inanimate processes has been pointed out. As discussed, adaptability, reproduction, and feedback mechanisms can also be accommodated within the proposed formalism, as mechanisms by which biological systems modify the survivability function to improve stability measures and fitness. While the large number of DOFs, stability measures, and noise sources make it difficult to model complex organisms and bio-networks, there is a clear potential application to relatively simple ones.  \\

The abundance of self-organised phenomena offers many opportunities to test this approach.  The procedure is to identify quasi-particles that are sensitive the SO, identify stability measures that depend of the quasi-particles' DOFs, and construct a survivability function that decreases with configurational stability. The dependence of the characteristics of the self-organised state on measurable parameters can also be included indirectly in the survivability function. For example, in the granular example, the inertial number can be controlled by shear rates, which affects the noise parameters. Similarly, the density should be correlated with the cell order distribution, $Q_k$. Including experimentally controlled parameters makes this formalism very flexible. The expectation values of characteristics of the long-term self-organised states, derived with the formalism can be then tested against observations. This author is looking forward to such tests.\\

\nin \underline{{\bf Acknowledgement}}: \\
Discussions with J.L. England and D.I. Goldman are gratefully acknowledged. \\

\nin \underline{{\bf Competing interests}}: \\
The author declares no competing interests. \\

\pagebreak

\begin{center} 
\begin{Large}{\bf  Supplemental material} \end{Large}
\end{center}



\noindent\underline{\textbf{The cell order and cell stress}}

\begin{figure}[!h]
 \includegraphics[width=0.6\linewidth]{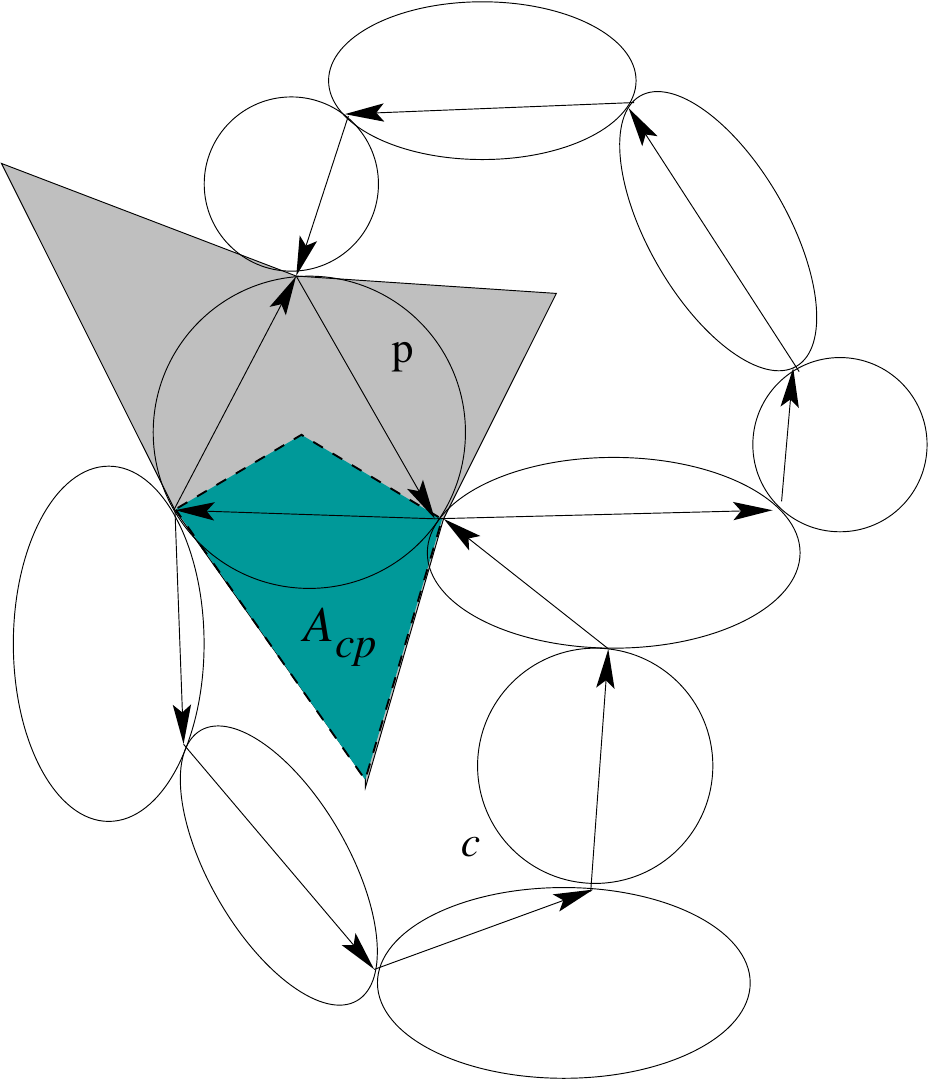}
 \caption{The structure around cell $c$. The quadron shared between cell $c$ and particle $p$ (shaded blue) is a quadrilateral whose diagonals are the line connecting the contacts of $p$, which border on $c$, and the line connecting the centroid of $p$ with the centroid of $c$. Its area is $A_{cp}$. The sum of these areas around $p$ is the area associated with particle $p$, $A_p$. The sum of these areas, which $c$ shares with the particles around it. 
The stress of cell $c$ is the weighted stresses of the particles around it, $\sigma_c=\sum_{p\in c}A_{cp}\sigma_p/A_p$.}
 \label{CellStress}
\end{figure}

A cell order is defined as the number of particles in contact enclosing. 
Each particle around the cell shares (generically) a quadrilateral with the cell, called 'quadron'~\cite{BaBl02,BlEd06,Bletal15}, shown in Fig.~\ref{CellStress}. The quadron is the most elementary area in the planar packing. The area associated with the cell is the sum of the areas of the quadrons it shares with its surrounding particles,
\beq
A_c = \sum\limits_{p\in c} A_{cp} \ .
\eeq
Similarly, the area associated with a particle is the sum of the areas of the quadrons that it shares with its surrounding cells,
\beq
A_p = \sum\limits_{c\in p} A_{cp} \ .
\eeq
The stress on a particle with $n$ force-carrying contacts is defined as 
\beq
\sigma_p = \frac{1}{A_p} \sum\limits_{j=1}^n \vec{f}_{jp}\otimes\vec{\rho}_{jp} \ ,
\eeq
where the sum runs over all the particles in contact with $p$, $\vec{f}_{jp}$ is the force that particle $j$ exerts on $p$ and $\vec{\rho}_{jp}$ is the position of the contact at which this force is applied. 
A cell stress is defined as the area-weighted sum of the stresses of the particles surrounding it, \\
\beq
\sigma_c = \frac{1}{A_c} \sum\limits_{p\in c} A_{cp} \sigma_p \ .
\eeq

\noindent\underline{\textbf{The independent cells partition function}}

Starting from the partition function in the main text, 
\begin{align}
Z=\prod\limits_{k=1}^K \left[\left(\sum_{c\in k}^{Q_kN_c)} e^{-F_c(k)} \right) e^{-\mu Q_kN_c}\right]  \ ,
\label{Z0}
\end{align}
rewrite it in terms of the joint conditional density of states of each $k$-cell subset
\begin{align}
Z=\Bigg\{&\prod\limits_{k=1}^K e^{-\mu} \int e^{-J_1f-J_2h-J_3(\theta-\phi)^2} \notag \\ 
&\times P\left[\{ X\}=\left(a,b,\theta,\sigma_{1},\sigma_{2},\phi |k\right)\right]d^6\left\{X\right\}  
\Bigg\}^{Q_k N_c} \ ,
\label{Z1}
\end{align}
$\theta$ and $\phi$ have been shown to be strongly correlated for stability~\cite{Jietal25}, with $\Delta\equiv \theta-\phi$ normally distributed around zero for all $k$,
\beq
P_{\Delta}(\Delta) = C e^{-\frac{\Delta^2}{2\delta^2}} \ .
\eeq
Otherwise, the degrees of freedom are assumed independent. 
Integrating over $\Delta$ first, one obtains
\begin{align}
Z=&\prod\limits_{k=1}^K \Bigg\{\frac{e^{-\mu}}{\sqrt{1+2J_3\delta^2}} \int e^{-J_1f-J_2h} \notag \\ 
&P_a(a|k) P_b(b|k) P_1(\sigma_1|k) P_2(\sigma_2|k) da\ db\ d\sigma_1\ d\sigma_2 
\Bigg\}^{Q_k N_c} \ .
\label{Z2}
\end{align}
The conditional PDF of $f(k)$, $P_f(f|k)$ can be obtained from those of $a$ and $b$ and, similarly, the PDF of $h$, $P_h(h|k)$ can be obtained from those of $\sigma_1$ and $\sigma_2$. Moreover, recent observations~\cite{Jietal25} show that by the scaling $h\to\hat{h}\equiv h/\bar{h}(k)$, where $\bar{h}(k)$ is the mean of $h$ over the conditional PDF $P_h(h|k)$, all the conditional PDFs collapse onto a single $k$=independent Weibull form, $W(\hat{h})$. 
For simplicity, assume that the conditional PDFs of $f$ also collapse onto a single form, $U[\hat{f}\equiv f/\bar{f}(k)]$. This means that the most probable values, $f_0(k)$ and $h_0(k)$, also collapse to one value each irrespective of $k$. Changing integration variables in (\ref{Z2}), and noting that sum over $Q_k$ is normalised, one obtains
\begin{align}
Z=& \left[\frac{ e^{-\mu}}{\sqrt{1+2J_3\delta^2}}\right]^{N_c} \times \notag \\ 
&\prod\limits_{k=1}^K \Bigg\{\int e^{-J_1\bar{f}(k)\hat{f}-J_2\bar{h}(k)(\hat{h}}  W(\hat{h}) U((\hat{f}) d\hat{f} d\hat{h} 
\Bigg\}^{Q_k N_c} \ ,
\label{Z3}
\end{align}
or
\begin{align}
Z= \left[\frac{ e^{-\mu}}{\sqrt{1+2J_3\delta^2}}\right]^{N_c} 
\prod\limits_{k=1}^K \langle e^{-J_1\bar{f}(k)}\rangle^{Q_k N_c} \langle e^{-J_2\bar{h}(k)}\rangle^{Q_k N_c} \ .
\label{Z4}
\end{align}
The result in the main text follows.\\

\noindent\underline{\bf A model for $W(\hat{h})$}

To facilitate explicit calculations, let us construct a model for the PDF of the cell stress stability measure, $W\left(\hat{h}\right)$.
Define the stress ratio of a cell of order $n$ as $h=\left(\sigma_1-\sigma_2\right)/\left(\sigma_1+\sigma_2\right)$, with $\sigma_1$ and $\sigma_2$ its principal stresses. The mean of this ratio over all the $Q_nN_c$ $n$-cells is 
\beq
\bar{h}(n) = \frac{1}{Q_n N_c} \sum\limits_{j\in n}^{Q_n N_c} h_j \ ,
\eeq 
where $Q_n$ and $N_c$ have been defined in the main text. The number of cells in this subset fluctuates during the steady state but, by the definition of the steady state, $\bar{h}(n)$ is constant in the limit $N_c\to\infty$. 
By design, the lower the value of $h_j$ the more stable cell $j$ is.  
Anticipating SO, define a normalised stress ratio of this subset of cells, $\hat{h}(n)\equiv h/\bar{h}(n)$, as well as the conditional PDF of this variable,
$p_n(\hat{h}\mid n)$, $0\leq\hat{h}(n)\leq\hat{h}_{max}(n)$. \\
During the dynamics, cells enter the $n$-order subset, via the cell dynamics described in~\cite{Waetal20}, each with a specific stress ratio, $h'$, and others leave it. Upon entering, an $n$-cell is nudged by the local stress. This either breaks the cell and it leaves the subset or causes it to rearrange to be more stable mechanically and $h'$ decreases, transitioning from $h'$ to $h<h'$. Let the transition rate be $f(h'\to h)$ and normalise $h$ and $h'$ as above. The conditional PDF must satisfy
\beq
p_n(\hat{h}\mid n) = \int\limits_{\hat{h}(n)}^{\hat{h}_{max}} p_n(\hat{h}'\mid n) f(\hat{h}'\to \hat{h}(n)) d\hat{h}' \ .
\label{int}
\eeq
Up to this stage, the model is general and, to derive a closed-form PDF, requires an explicit expression for the transition rates. 
As a simple illustrative example, consider the model $f(\hat{h}'\to \hat{h}(n))= g_0\hat{h}(n)$, with $g_0$ a numerical coefficient. Eq. (\ref{int}) can be solved explicitly now: 
\beq
p_n(\hat{h}\mid n) = g_0 \hat{h}(n) e^{-\int\limits^{\hat{h}(n)} g_0 x dx} = g_0\hat{h}(n) e^{-\frac{g_0}{2}\hat{h}(n)^2} \ .
\label{DifEq}
\eeq
As has been shown in~\cite{Jietal25}, one of the manifestations of the SO is that, under the above normalisation, all the conditional PDFs collapse onto a master form that is independent of $n$,
\beq
W(\hat{h}) =  g_0\hat{h} e^{-\frac{g_0}{2}\hat{h}^2} \ .
\label{DifEq}
\eeq
This PDF is almost exactly the one measured in simulations in~\cite{Jietal25}, supporting this mean field derivation. \\

\noindent\underline{\bf Calculations of $\langle \hat{h}\rangle$ and $\langle \hat{f}\rangle$ - mean field } 
\noindent\underline{\bf approximation}

Using the above PDF, one can calculate in this approximation the mean stress stability measure per cell,
\begin{align}
\langle \hat{h}\rangle = \prod\limits_{k=1}^K \left[\frac{I_2(k)}{I_1(k)}\right]^{Q_k} \ ,
\label{Zh1}
\end{align}
in which the $m$th moment of $\hat{h}$ is
\beq
I_m(k) \equiv g_0 \int\limits_0^{h_{max}} \hat{h}^m e^{-\frac{g_0}{2}\hat{h}^2 - J_2\bar{h}(k)\hat{h}} d\hat{h} \ .
\eeq
A straightforward, if cumbersome, integration yields
\beq
I_1(k) = u - v\chi \quad ; \quad I_2(k) = w -2u\chi + v\chi^2 \ ,
\eeq
where
\begin{align}
\chi \equiv & \frac{J_2 \bar{h}(k)}{g_0} \notag \\
v \equiv & \sqrt{\frac{\pi g_0}{2}}\chi e^{\frac{g_0}{2}\chi^2} 
\left\{\Phi\left[\sqrt{\frac{g_0}{2}}(h_{max}+\chi)\right]-\Phi\left(\sqrt{\frac{g_0}{2}}\chi\right)\right\} \notag \\
w \equiv & \left\{\gamma\left[\frac{3}{2},\frac{g_0}{2}\left(h_{max}+\chi\right)^2\right] - 
\gamma\left(\frac{3}{2},\frac{g_0}{2}\chi^2\right) \right\}e^{\frac{g_0}{2}\chi^2} \notag \\
u \equiv &1 - e^{-\frac{g_0}{2}h_{max}(h_{max}+2\chi)} \ ,
\label{Ij}
\end{align}
$\Phi$ is the error function, and $\gamma$ is the incomplete gamma function~\cite{GrRy07}. \\

Assuming, for illustration, that the PDF $U(\hat{f})$ can be derived in the same way as $W(\hat{h})$ and, therefore, that it has the same functional form, 
\beq
U(\hat{f}) =  g'_0\hat{f} e^{-\frac{g'_0}{2}\hat{f}^2} \ ,
\eeq
the calculation of $\langle \hat{f}\rangle$ proceeds along similar lines. 
\begin{align}
\langle \hat{f}\rangle = \prod\limits_{k=1}^K \left[\frac{I'_2(k)}{I'_1(k)}\right]^{Q_k} \ ,
\label{Zh2}
\end{align}
in which 
\beq
I'_m(k) \equiv g'_0 \int\limits_0^{f_{max}} \hat{f}^m e^{-\frac{g'_0}{2}\hat{f}^2 - J_1\bar{f}(k)\hat{f}} d\hat{f} 
\eeq
is the $m$th moment of $\hat{f}$. Equivalently to the above calculation, defining
\begin{align}
\chi' \equiv & \frac{J_1 \bar{f}(k)}{g'_0} \notag \\
v' \equiv & \sqrt{\frac{\pi g'_0}{2}}\chi' e^{\frac{g'_0}{2}\chi'^2} 
\left\{\Phi\left[\sqrt{\frac{g'_0}{2}}(f_{max}+\chi')\right]-\Phi\left(\sqrt{\frac{g'_0}{2}}\chi'\right)\right\} \notag \\
w' \equiv & \left\{\gamma\left[\frac{3}{2},\frac{g'_0}{2}\left(f_{max}+\chi'\right)^2\right] - 
\gamma\left(\frac{3}{2},\frac{g'_0}{2}\chi'^2\right) \right\}e^{\frac{g'_0}{2}\chi'^2} \notag \\
u' \equiv &1 - e^{-\frac{g'_0}{2}f_{max}(f_{max}+2\chi')} \ ,
\label{Ipj}
\end{align}
the integrations yields
\beq
I'_1(k) = u' - v'\chi' \quad ; \quad I'_2(k) = w' -2u'\chi' + v'\chi'^2 \ .
\eeq
Thus, 
\begin{align}
\langle \hat{h}\rangle &= \prod\limits_{k=1}^K \left[\frac{w -2u\chi + v\chi^2}{u - v\chi}\right]^{Q_k} \notag \\
\langle \hat{f}\rangle &= \prod\limits_{k=1}^K \left[\frac{w' -2u'\chi' + v'\chi'^2}{u' - v'\chi'}\right]^{Q_k}  \ .
\label{meanhf}
\end{align}
This leads to the equation of state mentioned  in the main text,
\beq
\langle \hat{f}\rangle = \prod\limits_{k=1}^K 
\left\{\left[\frac{u' - v'\chi'}{u - v\chi}\frac{w -2u\chi + v\chi^2}{w' -2u'\chi' + v'\chi'^2} \right]\right\}^{Q_k} \langle \hat{h}\rangle
\label{EoS}
\eeq

\noindent\underline{\bf Calculation of the mean laning speed - mean}
\noindent\underline{\bf field approximation} 

Underlying the formalism presented in the main text is the assumption that a self-organised state exists, namely that there formed already two lanes, moving at different speeds in the $y$-direction and the question is how do the deviations of individuals' velocities from this mode of propagation affect the characteristics of the overall phenomenon. 
The partition function for the laning SO phenomenon in the main text involves Gaussian integrals, but the coupling between the individuals' degrees of freedom makes it difficult to evaluate it exactly in general. As the aim is an illustration of the formalism, rather than delving deeply into a detailed analysis of this phenomenon, some simplifications are made for calculating the partition function proposed in the main text. 
Consider movement in the $y$-direction of $N$ individuals. Individual $q$ can travel at velocity $v_{qy}(\tan{\theta},1)$. The PDFs of $v_{qy}$ and $\zeta\equiv\tan{\theta}$ are assumed independent, with $P_v\left(v_{qy}\right)=\sum_{m=1}^M p_m\delta(v_{qy}-v_m)$ ($\sum_{m=1}^M p_m = 1$), $P_\zeta\left(\zeta_q\right)=1/(2\zeta_0)$, and $-\zeta_0<\zeta_q<\zeta_0$ (corresponding to $-\pi/2<\theta_q<\pi/2$ ).  Since each individual interacts only with $K$ nearest individuals, with $K\approx4$, it is convenient to use a version of the mean field approximation, i.e., that the nearest individuals move at a mean velocity $\vec{u}=(u_x,u_y)$. When $u_x\neq0$ there is a drift normally to the forward direction. The partition function in the main text reduces then to
\begin{align}
Z = &\int  e^{-\sum\limits_{q=1}^N \left[J\zeta^2 + \lambda K \left(\vec{v}_q-\vec{u}\right)^2 + \left(v_{qy}-u_{max}\right)^2\right]} \notag \\
&\times \prod\limits_{q=1}^N \Big\{P\left(v_{qy}\right)d v_{qy} d \frac{\zeta_q}{2\zeta_0}\Big\} \ ,
\end{align}
the integration over the $\delta$-functions of the speeds distribution is straightforward, leaving Gaussian integrals over the angle variables $\zeta_q$. These can be evaluated explicitly, resulting in
\begin{align}
Z = \left[\frac{\sqrt{\pi} e^{-\lambda K |u|^2}}{4\zeta_0}\sum\limits_{m=1}^M \frac{S_m}{\sqrt{\xi_m}}  e^{- \lambda K v_m\left(v_m-2u_y\right) + \xi_m C_m^2}\right]^N  
\end{align}
in which 
\begin{align}
\xi_m &\equiv J + \lambda K v_m^2 \\
C_m &\equiv \frac{\lambda K u_x v_m}{\xi_m} \\
S_m &\equiv \Phi\left[\sqrt{\xi_m}\left(\zeta_0 - C_m\right)\right] + \Phi\left[\sqrt{\xi_m}\left(\zeta_0 + C_m\right)\right]
\end{align}
and 
\begin{equation}
\Phi(w)=\frac{2}{\sqrt{\pi}}\int_0^w e^{-x^2}dx \notag
\end{equation} 
is the standard error function. \\

The mean speed in the $y$-direction is then 
\begin{align}
\langle v\rangle = \frac{\sum\limits_{m=1}^M \frac{S_m}{\sqrt{\xi_m}} v_m e^{- 2\lambda K v_m\left(v_m-u_y-u_{max}\right) + \xi_m C_m^2}}
{\sum\limits_{m=1}^M \frac{S_m}{\sqrt{\xi_m}}  e^{- 2\lambda K v_m\left(v_m-u_y-u_{max}\right) + \xi_m C_m^2}} \ .
\end{align}
This mean speed is plotted in Fig.~\ref{MeanSpeed2} in the main text as a function of $T\equiv 1/\lambda$, which is the analog of temperature in thermal systems, for $M=2$. In this plot the chosen parameters are: $K=4$, $\left(v_1, v_2, u_x, u_y, u_{max}\right)=\left(0.5, 1.5, 0., 1., 1.5\right)$ and $J = 0.1, 0.5, 0.9, 1.3, 1.7$. 
At very low noise, $T\to0$, the lane's speed is dominated by the fast individuals, but this speed decreases as the noise increases until, as $T\to\infty$, the mean speed is a weighted average of the two speeds, $\langle v\rangle\to p_1v_1+p_2v_2$. 
Notably, even in this simplistic model, a transition appears as $J$ is reduced, from a monotonic decrease of $\langle v\rangle$ with $T$ to an emergence of a minimum at some critical $J_c$. It has been checked that this behaviour and the transition persist for wider distributions with $M\geq 6$.

\end{document}